# Einstein and Boltzmann: Determinism and Probability

## or

## The Virial Expansion Revisited

E. G. D. Cohen

*The Rockefeller University, 1230 York Avenue, New York, New York 10065, USA*

**Abstract**    Boltzmann's Principle $S = k \ln W$ was repeatedly criticized by Einstein since it lacked a proper dynamical foundation in view of the thermal motion of the particles, out of which a physical system consists. This suggests, in particular, that the statistical mechanics of a system in thermal equilibrium should be based on dynamics. As an example, a dynamical derivation of the density expansions of the two-particle distribution function, as well as of the thermodynamic properties of a moderately dense gas in thermal equilibrium, is outlined here. This is a different derivation than the usual one based on Gibbs' probabilistic canonical ensemble, where dynamics is eliminated at the beginning and equilibrium statistical mechanics is reduced to statics. It is argued that the present derivation in this paper could, in principle, also be applied to other equilibrium properties and perhaps also to other fields.

I.  **INTRODUCTION**

In 1877 Boltzmann wrote a seminal paper: "On the relation between the Second Law of Thermodynamics and Probability Theory with respect to the laws of thermal equilibrium." [1]

In this paper Boltzmann made a new connection between three fundamental aspects of Thermodynamics and Statistical Mechanics of systems in equilibrium.

The relation Boltzmann proposed in the above paper was: $S = k \ln W$. Here, S is the entropy of a system in thermal equilibrium, W a probability, and k Boltzmann's constant.[1]

Einstein did not agree with the probabilistic formulation of this paper, which he called: "Boltzmann's Principle", and criticized this paper from 1905 until 1910.

I will give three examples of Einstein's "uneasiness" with the formulation of "Boltzmann's Principle". Before I do that, I will define what Boltzmann meant with "the probability W". It is the number of complexions, i.e. the number of microstates, which corresponds to a macrostate of a given total energy of a macroscopic system. If N is the number of (microscopic) particles in the system and $n_j$ of these particles have a (kinetic) energy $\varepsilon_j$, then $W = N! / \prod_j n_j$.[2]

---

[1] The relation $S = k \ln W$ cannot be found in the above cited paper, for more details see [2]. In addition, Boltzmann also discusses the applicability of this relation to systems not in equilibrium.

Einstein's first objection in 1905 was [3]: "The word probability is used in $S = k \ln W$ in a sense that does not conform to its definition as given in the theory of probability" since it is not normalized to one.[3]

Einstein's second objection in 1909 was [4]: "[My][4] point of view is characterized by the fact that one should introduce the probability [W] of a specific [macroscopic] state [of a system] in a *phenomenological* manner. In that way one has the advantage of not *interposing* any particular theory, for example, any Statistical Mechanics."[5]

Einstein's third objection in 1910, which is the origin of this paper, was: "Usually W is put equal to the number of complexions. [However,] in order to [actually] calculate W, one needs a *complete* [deterministic] *molecular-mechanical* theory of the system under consideration. Therefore, it is dubious whether the Boltzmann Principle has any meaning without such a theory or some other theory which describes the elementary processes.

---

[2] Boltzmann considered only ideal gases in equilibrium.
[3] Boltzmann called W a thermodynamic probability and mentions in this paper its normalization only in words.
[4] I have used in this paper Chapter 4 of Abraham Pais' book: "Subtle Is The Lord", which deals with Einstein's work in "Statistical Physics". Words between square brackets within quotations are Pais' or my inserts in a quoted text.
[5] This was done by Einstein by inverting Boltzmann's expression into $W = e^{S/k}$, where S is the phenomenological entropy of a system. Almost 50 years later, Onsager and Machlup [7] used the same relation for a system near (in local) equilibrium for a theory of fluctuations of systems in equilibrium.

$[S = k \ln W]$ seems [therefore] without content from a phenomenological point of view without giving in addition such a [deterministic] 'Elementary Theory.'"[6] [5]

I note that this remark of Einstein is applicable to *all* properties of a system in equilibrium and is then an alternative to Gibbs' probabilistic approach to Statistical Mechanics (1902) [6].

Therefore, the basic question arises: Is a dynamical derivation of the equilibrium properties of a classical system possible?

In the following I will discuss as a (non-trivial) example the derivation of the virial (density) expansions of the pair distribution function and the thermodynamic properties of a (classical) system in thermal equilibrium [9].

I will do this by presenting a formal virial expansion of the *non-equilibrium* pair distribution function in powers of the (number) density of the microscopic particles in a gas, which requires dynamics, and then reduce this expansion to the virial expansion of the pair distribution function of a system in thermal equilibrium. Similar virial expansions for the thermodynamic properties of a system in equilibrium follow from this.

---

[6] Professor Joel Lebowitz drew my attention to Einstein's Autobiography [8], where Einstein writes very positively about the crucial use Planck made of Boltzmann's Principle, which lead him to the correct law of heat-radiation. In my opinion there is no contradiction between Einstein's critical statement of 1910 and that of 1946. Einstein's earlier remarks refer to the absence of a proper foundation of $S = k \ln W$, not to its application. They can also be seen as a forerunner of his later criticism of the probabilistic nature of quantum mechanics, arguing that "God does not play dice."

## II. THE BASIC IDEA OF VIRIAL EXPANSIONS IN EQUILIBRIUM AND NON-EQUILIBRIUM

The virial or density expansions reduce the intractable $N\left(\sim 10^{23}\right)$-particle problem of a macroscopic gas in a volume V to a sum of an increasing number of tractable isolated few (1, 2, 3, …) particle problems, where each group of particles *moves* alone in the volume V of the system.

Density expansions will then appear, since the number of single particles, pairs of particles, triplets of particles, …, in the system are proportional to $n, n^2, n^3, ...,$ respectively, where n = N/V is the number density of the particles.

**Equilibrium**

Let me first present as background the virial expansion in equilibrium.

In a system of particles in equilibrium with *short-range*[7] interparticle interactions, the above mentioned procedure in equilibrium – which will be discussed in more detail below – leads to virial expansions of the pair distribution function $f_2^e(x_1, x_2; \beta)$ as well as of the thermodynamic quantities of an (equilibrium) system, such as, e.g., for the pressure $p^e(n,T)$.

---
[7] Short range means of the order of the size of the particles.

Here, the superscript e refers to an equilibrium system, T is the temperature of the system, and $\beta = \frac{1}{kT}$, where k is Boltzmann's constant. The pair distribution function $f_2^e(x_1, x_2; t)$ is the average number of particle pairs in a system in equilibrium.[8] The phases $x_1, x_2$ of the particles 1 and 2 refer to their positions and momenta, i.e. $x_i = \vec{q}_i, \vec{p}_i$ $(i=1,2)$, respectively. These virial expansions have been obtained before using Gibbs' probabilistic ensembles [6].

I will restrict myself in this paper to isolated groups of two- and three-particle contributions only. The same "procedure" holds for groups of $s > 3$ particles (cf [14]).

## III. THE VIRIAL EXPANSION OF THE PAIR DISTRIBUTION FUNCTION IN THERMAL EQUILIBRIUM FOR SHORT-RANGE INTER-PARTICLE INTERACTIONS

### A. Virial Expansion

For a system in equilibrium, the pair distribution function [9] $f_2^e(x_1, x_2; \beta) \equiv f_2^e(\vec{q}_1, \vec{p}_1, \vec{q}_2, \vec{p}_2; \beta)$ can be written as a *product* of $\vec{q}$ and $\vec{p}$ dependent functions:

---

[8] More precisely, the density of a pair of particles in their 12-dimensional phase space $(\vec{q}_1, \vec{q}_2, \vec{p}_1, \vec{p}_2)$.

$$f_2^e(x_1,x_2;\beta) \equiv f_2^e(\vec{q}_1,\vec{p}_1,\vec{q}_2,\vec{p}_2,\beta) = n_2^e(\vec{q}_1,\vec{q}_2;\beta) \bullet f_1^e(p_1;\beta) f_1^e(p_2;\beta). \quad (1)$$

Here,

$$f_1^e(p;\beta) = c e^{-\beta \frac{p^2}{2m}} \text{ with } p = |\vec{p}| \text{ and } c = (2\pi m k T)^{-3/2}, \quad (2)$$

the Maxwell velocity distribution, where m is the mass of a particle, and

$$n_2^e(\vec{q}_1,\vec{q}_2;\beta) = n^2 e^{-\beta \Phi_2(\vec{q}_1,\vec{q}_2)} + n^3 \int d\vec{q}_3 \left[ e^{-\beta \Phi_3(\vec{q}_1,\vec{q}_2,\vec{q}_3)} - e^{-\beta[\Phi_2(\vec{q}_1,\vec{q}_2)+\Phi_2(\vec{q}_1,\vec{q}_3)]} - \right.$$

$$\left. - e^{-\beta[\Phi_2(\vec{q}_1,\vec{q}_2)+\Phi_2(\vec{q}_2,\vec{q}_3)]} + e^{-\beta \Phi_2(\vec{q}_1,\vec{q}_2)} \right] + O(n^4), \quad (3)$$

where $\Phi_2(\vec{q}_1,\vec{q}_2)$ and $\Phi_3(\vec{q}_1,\vec{q}_2,\vec{q}_3)$ are the potential energies of isolated groups of two and three particles, respectively.

It will always be assumed in this paper that $\Phi_2(\vec{q}_1,\vec{q}_2)$, the potential energy between two particles 1 and 2, is: (1) repulsive, (2) spherically symmetric, (3) with a short range $\sigma$ of the order of the diameter of a particle and, (4) that for three or more interacting particles the interaction potential is additive, i.e. that e.g., $\Phi_3(\vec{q}_1,\vec{q}_2,\vec{q}_3) = \sum_{\substack{i<j \\ 1}}^{3} \varphi(r_{ij})$, where

$r_{ij} = |\vec{q}_i - \vec{q}_j|$ $(i,j=1,2,3)$, etc.

Here, $\varphi(r_{ij})$ is only $\neq 0$ when the two particles 1 and 2 overlap each other in space so that the distance between the two particles 1 and 2, $r_{12} \equiv |\vec{q}_1 - \vec{q}_2| \leq \sigma$, which will always be assumed to be the case in this paper[9].

The equilibrium power series expansion in n converges for sufficiently small n for all thermodynamic properties, as was shown by Ruelle in 1963 [11].

**B.  Cluster Property**

The *integrand* in Eq. (3) has the very important property that it is constructed in such a way that it vanishes for *separated configurations* of particles, i.e., whenever not *all* potential fields of the (e.g., 2, 3, …) particles overlap with each other in space. Therefore, only non-separated configurations where *all* three particles overlap 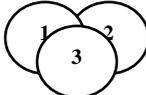 will contribute to the integral, while separated configurations 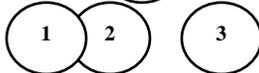 do not. Here, the diameter of the particles equals the interparticle potential range σ.

Thus, e.g., when in the integrand of Eq. (3) particle 3 does not overlap with both overlapping particles 1 and 2, the four exponents in the integrand all reduce to $\exp - \beta \Phi_2(\vec{q}_1, \vec{q}_2)$ so that it vanishes. (cf Eq. (13) and subsequent discussion below)

---

[9] If not, then $n_2^e(\vec{q}_1, \vec{q}_2) = n^2$.

This cluster property of the integrand defines the nature of the clusters considered in the virial expansion. The algorithm to obtain the expansion was originally proposed by H.D. Ursell in 1927 [10] and distinguishes the clusters in the virial expansions from those of other similarly named "cluster-expansions". Ursell formulated his algorithm for general potentials of three and more particles, without assuming additivity of intermolecular potentials. This is relevant if the interparticle interactions are not additive or for a quantum mechanical generalization of the virial expansions. [9]

**C.     Final Form of Equilibrium Virial Expansion**

For later I rewrite $f_2^e(x_1, x_2; \beta)$ using the cluster property of the integrand of Eq. (3), as well as Eqs. (1) – (3), in Hamiltonian form:

$$f_2^e(x_1, x_2; \beta) = n^2 e^{-\beta H_2(x_1, x_2)} + n^3 c \int' dx_3 e^{-\beta H_3(x_1, x_2, x_3)} + O(n^4), \tag{4}$$

where the Hamiltonians are $H_2(x_1, x_2) = \dfrac{p_1^2}{2m} + \dfrac{p_2^2}{2m} + \varphi(r_{12})$ and

$H_3(x_1, x_2, x_3) = \sum\limits_{i=1}^{3} \dfrac{p_i^2}{2m} + \sum\limits_{\substack{i<j \\ 1}}^{3} \varphi(r_{ij})$ and c normalizes the integration over the momentum $\vec{p}_3$ of particle 3 (cf Eq. (2)).

Here, the prime in Eq. (4) indicates that the integration over particle three is only for those configurations where *all* three particles 1, 2, and 3 overlap each other.

Eq. (4) follows from the Eqs. (1) -- (3) and the cluster property. I have also used here that the product of the two- and three-particle momenta distribution functions $f_1^e(p;\beta)$, when combined with the corresponding potential energy contributions – which due to the cluster property involve only genuine 3-particle overlaps – can be written in Hamiltonian form and that the integral in Eq. (3) can be rewritten as an integral over $x_3$ by multiplying with an integral $\int d\vec{p}_3 f_1^e(p_3;\beta) = 1$, which gives, with Eq. (2), the multiplication by c in Eq. (4).

## IV. <u>THE VIRIAL EXPANSION OF THE PAIR DISTRIBUTION FUNCTION IN NON-EQUILIBRIUM</u>

**A.** Non-equilibrium systems differ fundamentally from equilibrium systems in that dynamics *has* to be used in order to describe their physical properties. This is due to the presence of two features not present in equilibrium systems: non-vanishing currents (of heat, momentum, and particles) caused by the presence of gradients, e.g. of the thermodynamic quantities, such as the number density or the temperature. Furthermore, there is an explicit *time* dependence in addition to the space dependence in equilibrium. As a result there is then also *no* separation of $\vec{q}$ and $\vec{p}$ as in equilibrium since both are needed in dynamics.

For a (sufficiently) dilute system of particles not in thermal equilibrium, a virial expansion of the pair distribution function $f_2^{ne}(x_1,x_2;t)$ at a time t can be obtained by applying the same isolated small-group procedure as used in equilibrium. The superscript ne indicates a system not in equilibrium. This leads to a *formal* density expansion of $f_2^{ne}(x_1,x_2;t)$, given in the next section.

In non-equilibrium, the (static) overlaps of particles in equilibrium in space ("equilibrium collisions") are replaced by *genuine dynamical* collisions in space and time.

### B. Density expansion of the non-equilibrium pair distribution function

The density expansion of the non-equilibrium pair distribution can be written for short-range forces [15] as:

$$f_2^{ne}(x_1,x_2;t) = n^2 S_{-t}(x_1,x_2) f_1^{ne}(x_1;t) f_1^{ne}(x_2;t) +$$

$$+ n^3 \int dx_3 \left[ S_{-t}(x_1,x_2,x_3) - S_{-t}(x_1,x_2) S_{-t}(x_2,x_3) - \right.$$

$$\left. - S_{-t}(x_1,x_2) S_{-t}(x_2,x_3) + S_{-t}(x_1,x_2) \right] \bullet$$

$$\bullet f_1^{ne}(x_1;t) f_1^{ne}(x_2;t) f_1^{ne}(x_3;t) + O(n^4). \tag{5}$$

Here, it has been assumed that $f_2^{ne}(x_1,x_2;t)$ depends on the time only via $f_1^{ne}(x;t)$ [10]. This assumption, as well as the use of dynamical streaming operators $S_{-t}(x_1,...,x_s)$, was introduced by Bogolubov.[11] [12]. He obtained $f_2^{ne}(x_1,x_2;t)$ formally as a power series in the density n by an iterative solution of the BBGKY hierarchy [9,12], where the p-th approximation had to be obtained in terms of previous approximations < p so that the general term could not be written down explicitly, unlike in the cluster expansions used here.

The streaming operators $S_{-t}(x_1,...,x_s) \equiv e^{-t\mathcal{H}_s(x_1,...,x_s)}$ provide the solutions of the dynamical s-particle problem in giving the phases $x_i \equiv (\vec{q}_i, \vec{p}_i), i=1,...,s$ of an isolated group of s-particles at time $-t$ in terms

---

[10] This assumption was used by Bogolubov for very short times of the order of the duration of a binary collision. To the contrary, here it is relevant for sufficiently long times.
[11] The use of $S_{-t}$ operators instead of $S_t$ operators in Eq. (5) is due to the necessity to introduce a physically correct arrow of time in non-equilibrium statistical mechanics [14], which is absent in equilibrium.

of their initial phases at time $t = 0$. Thus, when acting on a function

$f(x_1,...,x_s)$, the $S_{-t}(x_1,...,x_s)$ replace the (initial) phases $x_1,...,x_s$ into those

at an earlier time $-t$ so that

$$S_{-t}(x_1,...,x_s)f(x_1,...,x_s) = f(S_{-t}(x_1,...,x_s)x_1,...,S_{-t}(x_1,...,x_s)x_s). \qquad (6)$$

The dynamical operators $\mathcal{H}_s(x_1,...,x_s)$ $(s = 1,2,3,...)$ occur in

Poisson's form of Hamilton's equations of motion for a system of s-particles

and solve the dynamical s-particle problem. They are defined by:

$$\mathcal{H}_s(x_1,...,x_s) = \sum_{i=1}^{s} \frac{\vec{p}_i}{m} \bullet \frac{\partial}{\partial \vec{q}_i} - \sum_{\substack{i<j \\ 1}}^{s} \theta_{ij} \qquad (7)$$

with

$$\theta_{ij} = \frac{\partial \varphi(r_{ij})}{\partial \vec{q}_i} \bullet \frac{\partial}{\partial \vec{p}_i} + \frac{\partial \varphi(r_{ij})}{\partial \vec{q}_j} \bullet \frac{\partial}{\partial \vec{p}_j}. \qquad (8)$$

Here, $\vec{p}_i \bullet \frac{\partial}{\partial \vec{q}_i}$ gives the rate of change of the position $\vec{q}_i$ of particle i and

$\theta_{ij}$ the rate of change of the momenta $\vec{p}_i$ and $\vec{p}_j$ of particles i and j due to the

interparticle potential (forces) per unit time during a binary, i.e. two-particle collision, respectively.

The combination of the four terms in the integrand of Eq. (5) have the

*same* cluster property as the Boltzmann factors had in the equilibrium case in the integrand of Eq. (3). Therefore, their contribution vanishes unless a genuine *dynamical* 3-particle collision occurs, where all three particles overlap each other during the collision at time t (cf Fig. 1d), so that no earlier contributions due to 2-particle collisions are counted and would then contribute again.

$$f_2^{ne}(x_1, x_2; t) = n^2 S_{-t}(x_1, x_2) f_1^{ne}(x_1; t) f_1^{ne}(x_2; t) +$$

$$+ n^3 \int dx_3 \left[ S_{-t}(x_1, x_2, x_3) - S_{-t}(x_1, x_2) S_{-t}(x_1, x_3) - S_{-t}(x_1, x_2) S_{-t}(x_2, x_3) + S_{-t}(x_1, x_2) \right] \cdot$$

$$\cdot f_1^{ne}(x_1; t) f_1^{ne}(x_2; t) f_1^{ne}(x_3; t). \qquad (9)$$

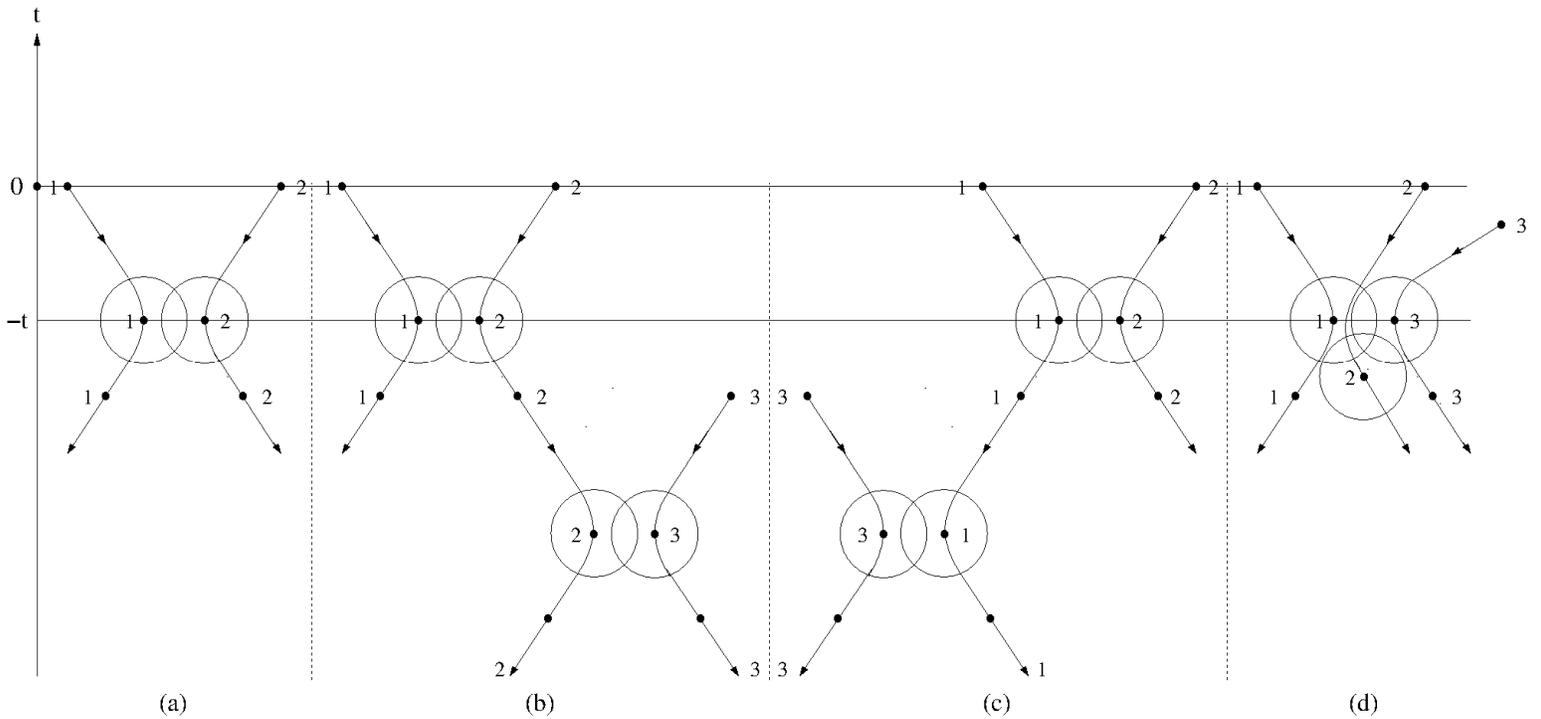

**FIGURE 1**    Two- and three-particle collisions which occur in the integral in Eq. (9), respectively.

(a) Binary collision of particles 1 and 2.
(b), (c) Two successive binary collisions of particles 1, 2, and 3.
(d) Genuine three-particle collision.

# V. DERIVATION OF THE EQUILIBRIUM DENSITY EXPANSION OF $f_2^e(x_1,x_2;\beta)$ FROM THE NON-EQUILIBRIUM EXPANSION OF $f_2^{ne}(x_1,x_2;t)$

## A. Non-equilibrium Virial Expansion

For comparison with the equilibrium expansions, Eqs. (1)-(4), we rewrite the density expansion of Eq. (9) of $f_2^{ne}(x_1,x_2;t)$ in the more explicit form:

$$f_2^{ne}(x_1,x_2;t) = n^2 e^{-t\mathcal{H}_2(x_1,x_2)} f_1^{ne}(x_1;t) f_1^{ne}(x_2;t) + n^3 \int dx_3 \Big[ e^{-t\mathcal{H}_3(x_1,x_2,x_3)}$$
$$- e^{-t\mathcal{H}_2(x_1,x_2)} e^{-t\mathcal{H}_2(x_1,x_3)} - e^{-t\mathcal{H}_2(x_1,x_2)} e^{-t\mathcal{H}_2(x_2,x_3)} + e^{-t\mathcal{H}_2(x_1,x_2)} \Big] \cdot$$
$$\cdot f_1^{ne}(x_1;t) f_1^{ne}(x_2;t) f_1^{ne}(x_3;t) + O(n^4), \qquad (10)$$

In the following two subsections, I will derive from this density expansion of $f_2^{ne}(x_1,x_2;t)$ the density expansion of $f_2^e(x_1,x_2;\beta)$.

## B. Equilibrium Virial Expansion

The static equilibrium virial expansion of $f_2^e(x_1,x_2;\beta)$ in Eqs. (1) – (3) can be obtained from the dynamical non-equilibrium density expansion of $f_2^{ne}(x_1,x_2;t)$ by replacing in Eq. (10) $f_1^{ne}(\vec{q},\vec{p};t)$ by $f_1^e(p;\beta)$, the Maxwell distribution function of the particles' velocities in equilibrium.

One uses the following correspondences:

| Nonequilibrium | Equilibrium |
|---|---|
| $f_1^{ne}(\vec{q}, \vec{p}; t)$ | $f_1^e(p; \beta)$ |
| $e^{-t \mathcal{H}_s(x_1, \ldots, x_s)}$ | $e^{-\beta H_s(x_1, \ldots, x_s)}$ |
| $t$ | $\beta = \dfrac{1}{kT}$ |

Since the integrand of the integral in Eq. (10) has the same cluster property as that in Eq. (3), the only contributions to this integrand are from genuine dynamical collisions, where all three particles overlap for the duration of their collision.

In fact, similar to the transition of Eq. (3) to Eq. (4), a transition can be made of Eq. (10) to Eq. (4) if $f_1^e(p;\beta)$ is substituted in Eq. (10) for $f_1^{ne}(x;t)$. The $f_2^{ne}(x_1, x_2; t)$ expansion in Eq. (10) becomes then [15]:

$$f_2^{ne}(x_1,x_2;t;\beta) = n^2 e^{-t\mathcal{H}_2(x_1,x_2)} f_1^e(p_1;\beta) f_1^e(p_2;\beta) +$$

$$+n^3 \int dx_3 \left[ e^{-t\mathcal{H}_3(x_1,x_2,x_3)} - e^{-t\mathcal{H}_2(x_1,x_2)} \bullet e^{-t\mathcal{H}_2(x_1,x_3)} - \right.$$

$$\left. -e^{-t\mathcal{H}_2(x_1,x_2)} e^{-t\mathcal{H}_2(x_2,x_3)} + e^{-t\mathcal{H}_2(x_1,x_2)} \right] f_1^e(p_1;\beta) f_1^e(p_2;\beta) f_1^e(p_3;\beta) + O(n^4). \quad (11)$$

The reduction of this hybrid dynamical non-equilibrium expansion, depending on t and $\beta$, to the equilibrium expansion (Eq. 4) will be carried out in the next session.

### C. Reduction of $f_2^{ne}(x_1,x_2;t;\beta)$ to $f_2^e(x_1,x_2;\beta)$

In spite of the term-by-term similarity of the composition of the terms in the integrand of the three-particle term of the non-equilibrium expansion Eq. (11) and the equilibrium expansion Eqs. (1) – (3), a term-by-term reduction of the former to the latter does not seem feasible. To the contrary, one first has to use the cluster property in order to obtain this reduction.

In the *two-particle term* on the right-hand side of Eq. (11), the two particles 1 and 2 are separated at t = 0 [12] since they only possess then kinetic and no potential energy. In order to contribute to the virial expansion, an overlapping configuration at time –t of their positions must occur and their momenta at $t = 0$ must therefore be such that they collide at time $-t$ in a (genuine) binary collision, which lasts for the duration of that collision, after which they separate.

---

[12] This means that they can be assumed to be uncorrelated, since the gas is very dilute in that they have not collided before, which is necessary for the proper arrow of time. (cf [14])

Similarly, in the *three-particle term* in Eq. (11), the three particles 1, 2, 3 are at $t = 0$ separated and must have then positions and momenta so that they collide at time $-t$ in a genuine three-particle collision in order to contribute to the three-particle integral.

Energy conservation for an isolated group of particles assures then that the original kinetic energy at $t = 0$ changes into varying potential and kinetic energy contributions during the genuine 2, 3, … particle collisions. The reduction of Eq. (11) is achieved in the following four steps:

1. The crucial observation is that, for isolated groups of $s = 2, 3, ...$ particles, the Hamilton-Poisson operators $\mathcal{H}_s(x_1, ..., x_s)$ in the streaming operators $S_{-t}(x_1, ..., x_s)$ acting on a product of s momentum distribution functions at time $t = 0$, as in Eqs. (5, 9), *conserve* the initial kinetic energy

$$\frac{p_1^2(0)}{2m} + ... + \frac{p_s^2(0)}{2m} \quad (s = 2, 3, ...) \text{ in the } \prod_{i=1}^{s} f_1^e(p_i; \beta) \text{ during their motion}$$

backwards over a time t.

This initial kinetic energy is then, during the collision at time -t, distributed into varying potential and kinetic energy contributions over the particles.

2. Thus, in the first term on the right-hand side of Eq. (11), the operator $e^{-t\mathcal{H}_2(x_1,x_2)}$ acts on the total (kinetic) energy of the particles 1 and 2, which are in a separated configuration at the initial time t = 0, since there is no potential energy. In order to obtain an overlapping configuration of the particles 1 and 2 at time $-t$ with $r_{12}(-t) \leq \sigma$, their positions and momenta at time 0 have to be such that a genuine binary collision occurs after the backward motion over a time t. As mentioned above, the initial kinetic energy will then be converted, during the genuine two-particle collision, into a sum of varying kinetic and potential energies, respectively, equal to the initial kinetic energy at t = 0. After the binary collision, the particles 1 and 2 will separate and regain their original kinetic energy, never to collide again.

3. Similarly, in the second term of Eq. (11), the streaming operator $e^{-t\mathcal{H}_3(x_1,x_2,x_3)}$ acts on a function of $x_1, x_2, x_3$, which are in a separated configuration. During the backward motion over a time t, their initial kinetic energy is transformed during a genuine three-particle collision into a sum of potential and kinetic energy which equals the initial kinetic energy.

In general, for s-particles, one has for the backward motion:

$$S_{-t}(x_1,...,x_s) \sum_{i=1}^{s} \frac{p_i^2(0)}{2m} \rightarrow \sum_{i=1}^{s} \frac{p_i^2(-t)}{2m} + \sum_{i<j}^{s} \varphi(r_{ij}(-t)) = H_s(q_1(-t),...,p_s(-t)). \qquad (12)$$

4. Then, with Eq. (12), and similar as in the equilibrium expansion, one can write $f_2^{ne}(x_1,x_2;t,\beta)$ in a form identical to the equilibrium expansion of Eq. (4):

$$f_2^e(x_1,x_2;\beta) = n^2 e^{-\beta H_2(x_1,x_2)} + n^3 c \int' dx_3 e^{-\beta H_3(x_1,x_2,x_3)} + O(n^4). \qquad (4)$$

The general term in this expansion for s-particles is:

$n^s c^{s-2} e^{-\beta H_s(x_1,...,x_s)}$, where the separation into q and p in equilibrium is regained in the Hamiltonian.

## C. **Thermodynamic Properties**

From the above result for $f_2^e(x_1,x_2;\beta)$, one can also obtain the virial expansions for the thermodynamic quantities of a system in equilibrium. As an example, the pressure $p^e(\beta)$ can be obtained by using Clausius' dynamical virial theorem[13] [9]:

---

[13] The same result can be obtained from an Ursell expansion of the partition function (cf Eq. (16)) of the system in equilibrium. It should be noticed that the standard derivation of the equilibrium density expansion is made in two steps: first, an expansion in terms of the fugacity z, followed by an expansion of z in terms of n [9]. This differs from the procedure in this paper, where only density expansions are considered.

$$p^e(\beta) = nkT\left[1 - \frac{1}{6}\int d\vec{r}_{12}\, n_2^e(r_{12};\beta)\, \vec{r}_{12} \bullet \frac{d\varphi(r_{12})}{d\vec{r}_{12}}\right]. \tag{13}$$

Using that $n^e_2(r_{12};\beta) = n^2 e^{-\beta\varphi(r_{12})} + O(n^3)$, one obtains

$$p^e(\beta) = nkT - \frac{2\pi}{3} n^2 \int \left[e^{-\beta\varphi(r_{12})} - 1\right] r_{12}^2 dr_{12} + O(n^3), \tag{14}$$

where the volume dependence of the pressure has been surpressed.

Here, the crucial role of the cluster property of the integrand in the second term on the right-hand side of Eq. (13) is clearly visible. This by noting that for any separated configuration of the particles 1 and 2 with $r_{12} > \sigma$, $\varphi(r_{12}) = 0$, so that the integral vanishes and only genuine two-particle overlaps contribute. Without the cluster property, the overwhelming contributions to the integral would come from *non*-overlapping configurations of the particles 1 and 2 in the entire volume V, which were already contained in the ideal gas contribution in the first term on the right-hand side of Eq. (13).

Similar expansions can be made for all other thermodynamic quantities [9].

## VI. **DISCUSSION**

1. The Eq. (4) is the most compact and physical representation of the equilibrium virial expansion, since each term contains only the contributions of s+1-particles above and beyond the contributions of all < s-particles, which makes the Hamiltonian form possible.

2. The non-equilibrium theory developed here has been formulated for spatially homogeneous non-equilibrium states of the system. Therefore, I have ignored the spatial *in*homogeneity of a system in a non-equilibrium state due to the presence of currents[14]. However, since we are only interested here in the transition from a non-equilibrium to an equilibrium state for sufficiently *long* times, this spatial inhomogeneity can be considered to be sufficiently small, for what Onsager calls an "aged" system on the verge of equilibrium [13], so that it can be ignored.[15] In this connection it is also relevant that energy conservation allows the time t in Eq. (11) to go to infinity.

3. Gibbs avoided all dynamics in his derivation of the properties of systems in equilibrium by noting that the basic Liouville Equation for the N-particle distribution function:

---

[14] This inhomogeneity was incorporated in $\mathscr{S}_t(x_1,...,x_s) = S_{-t}(x_1,...,x_s) \prod_{i=1}^{s} S_t(x_i)$ in [14].

[15] This is just a reformulation of the $0^{th}$ law of Thermodynamics, which states that an isolated system of particles in a non-equilibrium initial state will always go to an equilibrium state.

$$\frac{\partial f_N^{ne}(x_1,...,x_N;t)}{\partial t} = -\mathcal{H}_N(x_1,...,x_N) f_N^{ne}(x_1,...,x_N;t) \tag{15}$$

has as a solution, Gibbs' time-independent canonical ensemble:

$$f_N^e(x_1,...,x_N;\beta) = \frac{e^{-\beta H_N(x_1,...,x_N;\beta)}}{Z_N(\beta)}, \tag{16}$$

where $Z_N(\beta) = \int dx_1 ... \int dx_N e^{-\beta H_N(x_1,...,x_N)}$ is Gibbs' canonical partition function [6] and the volume dependence has been suppressed.

In this way he transformed the derivation of the thermodynamic properties of a system in equilibrium from a time-dependent dynamical to a time-independent static problem, thereby enormously simplifying the description of systems in equilibrium, since probabilities are much easier to handle than deterministic dynamics.

4. Phase transitions of systems in equilibrium have so far only been considered in the context of Gibbs' statistical mechanics. However, they can also be considered to be dynamical, since they can be triggered by a change in temperature, i.e. in the mean *kinetic* energy of the molecules of a system in equilibrium. A dynamical derivation of them would, in my opinion, be very instructive.

5. It seems worthwhile to me to emphasize the cluster property of Ursell's expansion. This property assures – in the absence of long-range interactions in space – that the behavior of a not-too-dense system can be described by considering sequences of isolated groups of two-, three-, etc. particles.

One could imagine that similar expansions might be useful in a variety of problems not necessarily restricted to Statistical Mechanics or even physics. In fact, the Ursell expansion of the partition function for a system in equilibrium [9] is an application in physics of the inclusion-exclusion principle in set theory.[16]

6. In view of the dynamical derivation of the virial expansions in equilibrium given here, and the usual one based on Gibbs' probabilistic canonical ensemble, I quote a remark of Ehrenfest: "When a result can be obtained in two different ways, one stands on two legs rather than on one."

It seems to me that the Hamiltonian formulation of the virial expansion, obtained in this paper by a dynamical derivation when compared with Gibbs' static derivation, is an example of Ehrenfest's dictum.

---

[16] I am indebted to Professor Joel Cohen for pointing this out to me.